\documentclass{article}
\usepackage[T1]{fontenc} 
\usepackage[utf8]{inputenc} 
\usepackage{ismir,amsmath,cite,url}
\usepackage{graphicx}
\usepackage{color}

\usepackage{multirow}
\usepackage{booktabs}
\usepackage{enumitem}

\usepackage{lineno}

\title{PiCoGen2: Piano cover generation with transfer learning approach and weakly aligned data}

\multauthor
{Chih-Pin Tan$^{1,2}$ \hspace{3mm} Hsin	Ai$^1$  \hspace{3mm} Yi-Hsin	Chang$^1$  \hspace{3mm} Shuen-Huei Guan$^2$  \hspace{3mm} Yi-Hsuan  Yang$^1$} {
 $^1$ Graduate Institute of Communication Engineering, National Taiwan University, Taiwan\\
$^2$ KKCompany Techonologies\\
{\tt\small tanchihpin0517@gmail.com, yhyangtw@ntu.edu.tw}
}

\def\authorname{C.-P. Tan, H. Ai, Y.-H. Chang, S.-H. Guan, and Y.-H. Yang}

\def\fnmap{F_\text{time}}
\def\fnpair{F_\text{beat}}

\begin{document}

\maketitle
\begin{abstract}
Piano cover generation aims to create a piano cover from a pop song. Existing approaches mainly employ supervised learning and the training demands strongly-aligned and paired song-to-piano data, which is built by remapping piano notes to song audio. This would, however, result in the loss of piano information and accordingly cause inconsistencies between the original and remapped piano versions.
To overcome this limitation, we propose a transfer learning approach that pre-trains our model on piano-only data and fine-tunes it on weakly-aligned paired data constructed without note remapping.
During pre-training, to guide the model to learn piano composition concepts instead of merely transcribing audio, we use an existing lead sheet transcription model as the encoder to extract high-level features from the piano recordings.
The pre-trained model is then fine-tuned on the paired song-piano data to transfer the learned composition knowledge to the pop song domain. Our evaluation shows that this training strategy enables our model, named PiCoGen2, to attain high-quality results, outperforming baselines on both objective and subjective metrics across five pop genres.

\end{abstract}
\section{Introduction}\label{sec:introduction}

\begin{figure}
    \centerline{{
    \includegraphics[width=0.9\columnwidth]{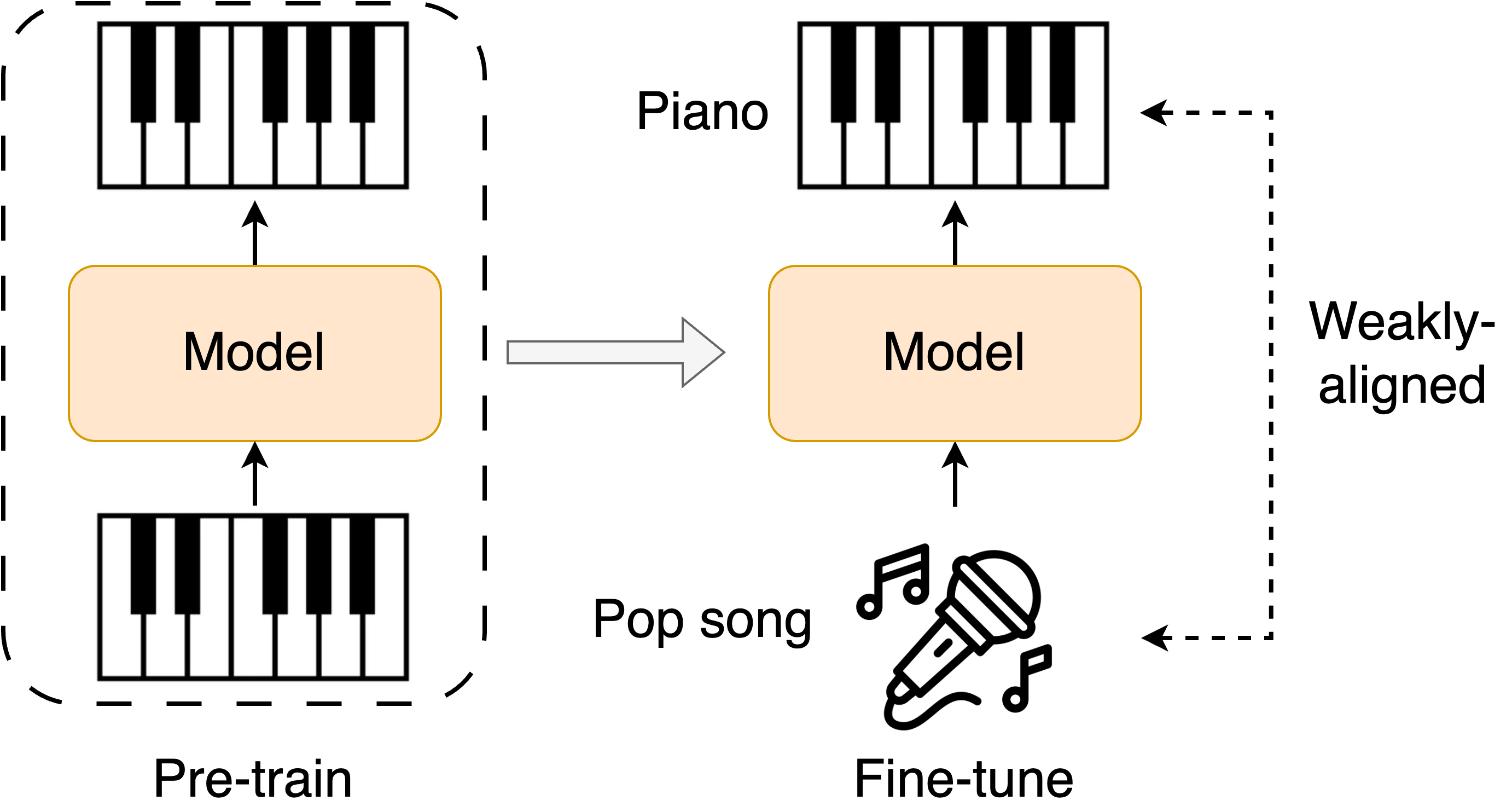}}}
    \caption{The proposed model is trained with two stages: firstly pre-trained on piano-only data and then fine-tuned on the weakly-aligned song-to-piano pairs.}
    \label{fig:cover}
\end{figure}

Piano cover generation, which involves recreating or arranging an existing music piece as a new piano version, is popular within music-creative communities and the music production industry. On media sharing sites like YouTube, piano cover creators often have lots of subscribers. Additionally, many music producers create and distribute piano arrangements on music streaming platforms.

Attempts have been made in the field of music information retrieval (MIR) to automatically generate piano covers from existing musical pieces.
Takamori \emph{et al.} \cite{takamori2019audio} proposed a regression method to generate piano reductions, which can be considered simplified versions of piano covers, using acoustic features and structural analysis of the input music.
With the recent surge in deep learning, Choi \emph{et al.} \cite{choi2023pop2piano} introduced a model named Pop2Piano that tackles piano cover generation by leveraging the concept of piano transcription and employing the MT3 architecture \cite{gardner2021mt3}, originally designed for transcription, as their model backbone.
They collected pop songs and the corresponding piano covers from the Internet, and built a song-piano \emph{synchronized} dataset by ``remapping'' the piano notes to the song audio with a warping algorithm (thereby modifies, or warps, the piano cover).
The algorithm entails evaluating the similarity between the pitch contour of the vocal signal extracted from the song audio with the top line of the piano MIDI.
They then trained the model with the synchronized data, guiding the model to learn the pitch and onset/offset timing 
of each note 
in the generated piano cover.

However, as shown in \tabref{tab:duration_tempo}, the statistics in the ratio of audio length difference and tempo difference between the original songs and original piano covers (i.e., before note-remapping) they collected\footnote{\url{https://github.com/sweetcocoa/pop2piano/blob/main/train_dataset.csv}} 
show that a piano cover and its original song are not perfectly aligned to each other (for otherwise the difference ratio would be equal to 1.00).
This indicates that the tasks cover generation and transcription are inherently different,
and that forcing a piano cover to be synchronized with its original song may be inappropriate.
Actually, we notice that the note-remapping process of Pop2Piano---i.e., adjusting piano note timing according to the time mapping function obtained by synchronizing piano notes to the song audio---breaks the relation of original piano notes and thereby incurs loss of piano information. Moreover, from a musical perspective,
the way human creates piano covers 
is by nature different from 
the way human transcribes music. 
For cover generation, musicians may firstly analyze the original song in terms of aspects such as melody, chord progression and rhythm section,  then decide how to interpret the original song with their composition knowledge, and finally make the piano cover based on the piano performance techniques.

Inspired by the process of human composition for piano cover songs, we propose 
in this paper a novel approach for piano cover generation by involving the concept of transfer learning\cite{zhuang2020comprehensive}.
Instead of relying on the \emph{strongly-aligned} pairs\cite{zalkow2020using}
that necessitates note-remapping,
we use \emph{weakly-aligned} data with the correspondence in ``beat'' level between song-piano pairs.
This approach incurs no rhythmic distortion of the piano covers, retaining their musical quality.
Besides, to mitigate the inaccuracy of data alignment, the model is pre-trained on piano-only data to learn the concept of piano performance first, and then fine-tuned on the weakly-aligned paired data to learn the conversion of song to piano, as shown in Figure \ref{fig:cover}.
We also employ a prior model SheetSage \cite{donahue2022melody}, pre-trained for lead sheet transcription, as an encoder component that helps our model learn high-level musical concepts for cover generation.

We compare the proposed model, named ``PiCoGen2'', against other baselines with objective and subjective measures, validating the effectiveness of the weak-alignment method for pairing and the two-step training strategy.
We share source code and audio samples at a project page.\footnote{\url{https://tanchihpin0517.github.io/PiCoGen/}}

\begin{table}
    \begin{center}
    \begin{tabular}{c|c|c}
     \toprule 
     duration deviation & tempo deviation & IOI deviation \\
     \midrule
     1.10 $\pm$ 0.12 & 1.16 $\pm$ 0.25 & 1.14 $\pm$ 0.17 \\
    \bottomrule 
    \end{tabular}
    \end{center}
    
    \caption{
    The first two statistics contrast the original songs with their original piano covers (i.e., no note-remapping) in the Pop2Piano dataset \cite{choi2023pop2piano}, evaluating the length of the \textbf{duration} (in seconds) of the longer one divided by that of the shorter one, and similarly the deviation ratio in \textbf{BPM}.  The last statistic is similarly the deviation ratio in terms of the average inter-onset intervals (\textbf{IOIs}; in seconds), but between the original \& adjusted (synchronized) piano covers.
    }
    \label{tab:duration_tempo}
\end{table}

\section{Background}\label{sec:background}

Piano arrangement, i.e., the process of reconstructing and reconceptualizing a piece, is related to various conditional music generation tasks, including lead sheet\footnote{A music notation consisting of lead melody and chord progression.}-conditioned accompaniment generation, transcription and reorchestration,
and piano reduction \cite{elowsson2012algorithmic, nakamura2015automatic, wang2020pop909}.
Beyond arrangement, piano cover generation involves creating new musical elements and modifying the original elements via improvisation, tempo changes, stylistic shifts, etc.
We briefly review some related topics below.

\textit{Symbolic-domain music generation} is about generating music in a symbolic form such as pianorolls \cite{musegan} and discrete MIDI- (Musical Instrument Digital Interface) \cite{musictransformer} or REMI-like tokens \cite{huang2020pop,hsiao2021compound,huang24ismir,viettoan24arxiv}, rather than audio signals.
The task encompasses unconditional generation (i.e., from-scratch generation) and conditional generation.
While the goal of piano cover generation is to generate piano audio given a song audio input, we can treat it as a conditional symbolic music generation task, 
for we can generate piano in the MIDI domain first, and then use 
off-the-shelf high-quality piano synthesizers to convert it into audio.


\textit{Automatic music transcription} (AMT), which aims to precisely transcribe music content from audio signals into a symbolic representation over time, is also related to piano cover generation.
AMT tasks can be categorized based on the completeness of information captured from the input audio.
One category of AMT tasks aims to capture all music content presenting in the audio, such as automatic piano transcription \cite{gardner2021mt3, hawthorne2017onsets, benetos2019amt, toyama2023apt, hawthorne2021sequence, kong2021high}.
These methods transcribe the complete polyphonic piano performance from the audio signal.
Another category focuses on transcribing a reduced representation of the input, like melody transcription\cite{paiva2005auditory, paiva2005detection} and lead sheet transcription\cite{matti2008auto, weil2009automatic, donahue2022melody}.
These tasks extract only the lead melody line and chord progressions, representing a sparse subset of the full musical content.
Piano cover generation also requires the exploration of music content reduction and additionally relies on generative modeling conditioned on the reduced representation.
For example, Pop2Piano uses MT3\cite{gardner2021mt3}
as its backbone to convert audio features into a symbolic piano performance representation.
However, following the paradigm of transcription approaches, Pop2Piano requires paired data consisting of pop songs and their corresponding temporally-synchronized piano cover.

\textit{Transfer learning} is generally consider as the concept of adopting the model to the target domain by re-using parameters that are trained on a source domain, thereby transferring the knowledge between the domains\cite{weiss2016survey}.
There have been several works on transfer learning in the field of MIR, e.g.,
music classification\cite{hamel2013transfer, van2014transfer, choi2017transfer} and
music recommendation\cite{choi2016towards, liang2015content}.
However, to our best knowledge, little attempts have been made to apply transfer learning to the task of cover song generation.

Besides Pop2Piano \cite{choi2023pop2piano}, this work is also closely related to PiCoGen \cite{tan2024picogen}, an early version of the current work. We explore the two-stage training strategy for piano cover generation for the first time there. However, in PiCoGen we use discrete symbolic lead sheet as the intermediate representation, instead of continuous conditions supplied by an encoder as done here (see Section \ref{sec:method:modl}). 
We note that the sampling process of lead sheet extraction in PiCoGen might loss musical information such as instrumentation and vibes of the input audio. Moreover, we do not explore the idea of transfer learning (Section \ref{sec:method:trans}) there.\footnote{As the previous work \cite{tan2024picogen} was also under review at the time we submitted the current paper, we did not empirically compare PiCoGen and PiCoGen2 in the experiments here. Instead, we provide examples of their generation results for the same input songs on the demo page, which should demonstrate that PiCoGen2 works better than PiCoGen.}

\begin{figure*}[t]
    \centerline{{
    \includegraphics[width=2.0\columnwidth]{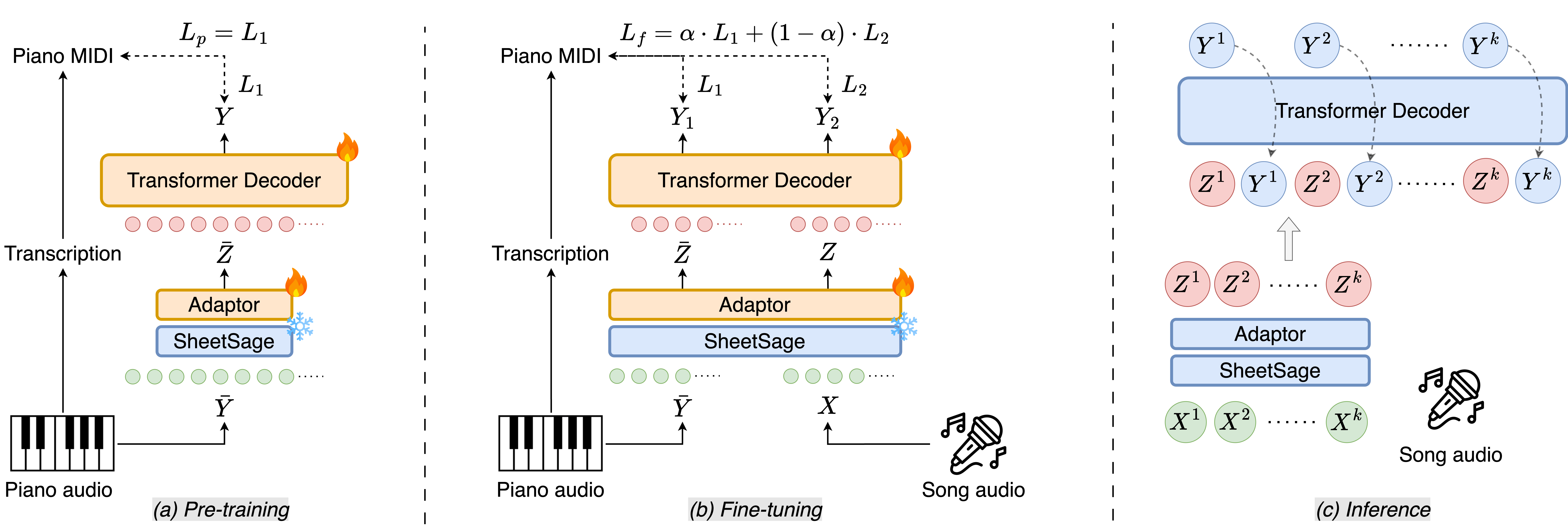}}}
    \caption{A diagram of the proposed model, PiCoGen2. The fire and snowflake symbols indicate the trainable and frozen parts. For example, the parameters for SheetSage \cite{donahue2022melody}, a model pre-trained for lead sheet transcription, are always frozen.
    }
    \label{fig:model}
\end{figure*}

The work of Wang \emph{et al.}\cite{wang2022audio} is also  related, for they deal with the similar problem of converting audio signals into piano MIDI performances. However, they apply a  piano transcription prior and thus using strongly-aligned data as Pop2Piano \cite{choi2023pop2piano}, and they employ a more sophisticated disentanglement-based method to get an intermediate representation. 
Moreover, they assume that the vocal  of the input audio has been removed beforehand, thus actually generating a piano backing track rather than a piano cover.










\section{Methodology}\label{sec:methodology}

Viewing piano cover generation as a conditional symbolic music generation task, we formulate it as a sequence-to-sequence problem.
The objective is to generate a sequence of symbolic tokens $Y$ representing the piano performance, conditioned on the input audio $X$ of the original song.

\subsection{Weakly-Aligned Data}

In Pop2Piano, Choi \emph{et al.}  \cite{choi2023pop2piano} propose a data preprocessing algorithm to synchronize the piano MIDI to the song audio. They utilize SyncToolBox \cite{muller2021sync} to analyze the chroma features of two audio segments to obtain a warping path of mapping the time from the piano performance to the song audio.
Based on the analysis, they adjust the timing of notes transcribed from the piano performance by using a linear mapping function calculated from the temporal warping information.
These remapped notes is then quantized to align with the beat locations of the song audio.
However, 
the rhythmic distortion caused by note-remapping is practically unavoidable, even disregarding the inaccuracy of the synchronization process.
The chroma feature only reflects a rough overall alignment between the piano performance and song audio which cannot precisely describe the nuanced amount of timing shift for each individual note.  
This is evident when examining the changes in the inter-onset intervals (IOIs) between the original piano notes and the remapped version, shown in \tabref{tab:duration_tempo}.


To avoid the rhythmic distortion of note-remapping, we propose a weak-alignment approach that does not change the timing of piano notes.
The idea is to let the alignment rely on only the  \emph{beats} of each song-to-piano pair.
We construct the time mapping function $\fnmap$ by computing the warping path for the audio pair like the way of Pop2Piano. 
Given a time of piano performance $t_p$, the function outputs the corresponding time of song audio $t_s = \fnmap(t_p)$ according to the temporal warping information.
Specifically, we detect the beat locations with Beat Transformer\cite{zhao2022beat} to get the beat times $Q^{p} = [q^{p}_{1}, \dots, q^{p}_{l_p}]$ of the piano performance and $Q^{s} = [q^{s}_{1}, \dots, q^{s}_{l_s}]$ of the song audio, where $l_p$ and $l_s$ denote the number of beats of each of them.
Then we define an aligning function $\fnpair$ as:
\begin{equation}
\fnpair(i) = \mathop{\arg\min}_{j}(\fnmap(q^{p}_{i}) - q^{s}_{j}). 
\end{equation}
For any beat index $i \in [1, ..., l_p]$ of the piano performance, the aligning function outputs the corresponding beat index $j \in [1, ..., l_s]$ of the song audio, and $q^{s}_{j}$ is the nearest beat time to $\fnmap(q^{p}_{i})$.
We consider a song-piano pair to be weakly-aligned if the correspondence between them is determined by $\fnpair$.
See the project page for an illustration.

\subsection{Model}
\label{sec:method:modl}

An aerial view of our model is depicted in \figref{fig:model}.
We employ a decoder-only Transformer to accept an input sequence bundling condition $X$ (song audio) and target $Y$ (piano performance) together, and generates the output tokens for $Y$ autoregressively.
This approach of providing both the condition and target as a bundled input sequence to the Transformer has been applied in previous studies\cite{hsiao2021compound, wu2023compembellish,huang24ismir} and has shown success in better informing the model of the temporal correspondence between the condition and desired output.
We divide $Y$ into \emph{bars} with the detected beat information and get $Y = [Y^1, \dots, Y^{B_p}]$, where $B_p$ is the number of bars in the piano cover,
and there exists an song audio sequence $X = [X^1, \dots, X^{B_p}]$ for $Y$, where each sub-sequence $X^k$ is weakly aligned to $Y^k$.  
We then rearrange them with an interleaving form
and train the decoder with the bar-wise mix $S = [X^1, Y^1 \dots, X^{B_p}, Y^{B_p}]$.
The decoder model would learn to generate $k$-th bar of piano performance $Y^{k}$ depending on (i.e., can attend to) the current and preceding sub-sequences of song audio $[X^1, \dots, X^{k}]$ and the preceding sub-sequences of piano performance $[Y^1, \dots, Y^{k-1}]$.

To reduce the sequence length of $X$ and extract better musical information,
we employ a prior audio encoder to transform $X$ into an intermediate representation $Z$.
Different from those works 
which use Mel-spectrograms \cite{gardner2021mt3} or audio codecs \cite{copet2023simple} for $Z$,
we use SheetSage\cite{donahue2022melody}, which is trained for  lead sheet transcription, cascaded with an neural adapter as the prior audio encoder.
We consider the output embeddings of SheetSage
more suitable for representing the input, since they carry information of musical elements connecting a  cover with the original song, such as melody, chords and vibes.
With the prior encoder, the song audio $[X^1, \dots, X^{B_p}]$ is transformed into a sequence of latent embeddings $[Z^1, \dots, Z^{B_p}]$ before being passed to the decoder,
yielding the input sequence $[Z^1, Y^1 \dots, Z^{B_p}, Y^{B_p}]$ of the decoder, as illustrated in \figref{fig:model}c.

\subsection{Transfer Learning}
\label{sec:method:trans}



While the weak-alignment approach eliminates inner temporal distortions for piano performance, there can still be alignment errors between the piano segments and their corresponding song segments.
This is because a piano cover is not guaranteed, in the beat level, to have a strict one-to-one mapping with the original song.


To abate such alignment errors, we propose a transfer learning-based training strategy, dividing the training into two steps: pre-training (\figref{fig:model}a) and fine-tuning (\figref{fig:model}b).
In the \emph{pre-training} stage, we train the model with an input sequence $\bar{S} = [\bar{Y}^{1}, Y^1, \dots, \bar{Y}^{B_p}, Y^{B_p}]$ 
where $\bar{Y}$ is the original piano audio recording of the symbolic piano tokens $Y$.
The same as the song audio, the original recording $\bar{Y}$ is encoded to an inter-representation $\bar{Z}$ by the prior encoder.
We expect the model to learn to generate piano performances $Y$ with high-level musical features extracted by SheetSage from the piano audio $\bar{Y}$, rather than merely detecting note onsets/offsets like in a piano transcription task.
Importantly, there will be no alignment errors between $Y$ and $\bar{Y}$, ensuring that 
the model can firstly learn the complete concept of piano composition and generation in the pre-training stage, without being impeded by cross-domain alignment issues.

In the \emph{fine-tuning} stage, we train the model with the mixture of $\bar{S}$ and $S$.
Following \cite{wu2020jazz, wu2021musemorphose, copet2023simple}, we train the model with the objective of minimizing the cross entropy loss on the tokens of piano performance $Y$.
Let $L_1$ and $L_2$ stand for the cross entropy losses for $\bar{S}$ and $S$, respectively. The loss $L_{p}$ in the pre-training stage and the loss $L_{f}$ in the fine-tuning stage can be writeen as:
\begin{equation}
\begin{aligned}
L_{p} &= L_1 \,,\\
L_{f} &= \alpha \cdot L_1 + (1-\alpha)\cdot L_2 \,,
\end{aligned}
\end{equation}
where $\alpha$ is  the weighting factor  determining the proportion of losses contributed from $\bar{S}$ and $S$ during fine-tuning.
We expect that $\alpha$ helps the model retain the knowledge about piano performance learned from the pre-training stage.

\subsection{Data Representation}
For the piano performance sequences $Y$, we adopt a modified version of the REMI token representation\cite{huang2020pop}, which has been shown to work well for modeling pop piano.
Our representation consists of 7 token classes.
\texttt{Spec} contains special tokens such as \texttt{[bos]} (beginning-of-sentence) and \texttt{[ss]} (song-start) for controlling the model behavior.
\texttt{Bar} indicates the property of each bars.
\texttt{Position}, \texttt{Chord} and \texttt{Tempo} are metric-related tokens for 16th-note offsets within bars, chord changes (11 roots $\times$ 12 qualities), and tempo changes (64 levels).
\texttt{Pitch}, \texttt{Duration} and \texttt{Velocity} are note-related tokens for note pitches (A0 to C8), durations (1 to 32 16th-notes), and note velocities (32 levels).
There are in total 428 tokens in the vocabulary.
In our implementation, \texttt{[Bar\_start]} and \texttt{[Bar\_end]} always occur at the start and end of each bar in the input sequence $S$ and $\bar{S}$.

\section{Evaluation}

\subsection{Dataset}\label{db}
We follow the instructions provided in the Pop2Piano source code to rebuild the training dataset, collecting 5,844 pairs of pop songs and their corresponding piano covers from the Internet.
We filter out song pairs with a melody chroma accuracy (MCA)\cite{raffel2014mir_eval} lower than 0.05 or an audio length difference exceeding 15\%, leaving 5,503 remaining pairs.
In the pre-training stage, all the piano performances from these remaining pairs are used for training.
In the fine-tuning stage, we remove invalid bars from the piano performances where the first and last beats of a bar were mapped to the same beat of the original song by the mapping function $\fnmap$.
Around 50\% of the bars are removed from the piano performances accordingly. 
We note that the large number of such invalid bars implies the alignment algorithm of Pop2Piano \cite{choi2023pop2piano} may not be robust enough and future work can be done to study  this.

For objective and subjective evaluations, we collect additional 95 song-to-piano pairs from the Internet, containing 19 Chinese Pop (\textbf{Cpop}), 20 Korean Pop (\textbf{Kpop}), 16 Japanese Pop (\textbf{Jpop}), 20 Anime Song (\textbf{Anime}),  20 Western Pop (\textbf{Western}) pairs.
All the songs contain vocals. 
We share the URLs of these songs at the project page.

\def\ptpobj{\textbf{0.42} $\pm$ 0.07 & 0.86 $\pm$ 0.09 & 2.46 $\pm$ 0.18}
\def\atmobj{0.15 $\pm$ 0.03 & 0.77 $\pm$ 0.06 & 2.41 $\pm$ 0.17}
\def\ourobj{0.17 $\pm$ 0.06 & 0.84 $\pm$ 0.06 & 2.46 $\pm$ 0.22}
\def\wopreobj{0.16 $\pm$ 0.05 & \textbf{0.87} $\pm$ 0.06 & \textbf{2.45} $\pm$ 0.23}
\def\wofineobj{0.15 $\pm$ 0.05 & 0.81 $\pm$ 0.06 & 2.57 $\pm$ 0.19}
\def\transobj{0.19 $\pm$ 0.06 & 0.67 $\pm$ 0.09 & 2.78 $\pm$ 0.30}
\def\humanobj{0.16 $\pm$ 0.06 & 0.81 $\pm$ 0.06 & 2.59 $\pm$ 0.18}

\def\ptpsub{2.71 $\pm$ 0.98 & 2.63 $\pm$ 1.01 & 2.72 $\pm$ 1.1}
\def\atmsub{- & - & -}
\def\oursub{\textbf{3.48} $\pm$ 0.93 & \textbf{3.55} $\pm$ 1.06 & \textbf{3.66} $\pm$ 1.02}
\def\wopresub{3.09 $\pm$ 1.03 & 2.96 $\pm$ 1.02 & 3.22 $\pm$ 1.09}
\def\wofinesub{3.09 $\pm$ 1.02 & 3.30 $\pm$ 1.07 & 3.08 $\pm$ 1.16}
\def\transsub{1.48 $\pm$ 0.74 & 1.69 $\pm$ 0.88 & 1.45 $\pm$ 0.71}
\def\humansub{4.30 $\pm$ 0.87 & 4.23 $\pm$ 0.95 & 4.33 $\pm$ 0.9}

\begin{table*}
    \centering 
    \resizebox{2.0\columnwidth}{!}{
    \begin{tabular}{ l|ccc|ccc }
        \toprule 
        \multirow{2}{*}{Model} & \multicolumn{3}{|c}{\emph{objective evaluation}} & \multicolumn{3}{|c}{ \emph{subjective evaluation} ($\in [1,5]$)}  \\ 
        & $MCA$\,$\uparrow$ & $GS$\,$\uparrow$ & $H_4$\,$\downarrow$ & OVL\,$\uparrow$ & SI\,$\uparrow$ & FL\,$\uparrow$ \\
        \midrule
        Pop2Piano\cite{choi2023pop2piano} & \ptpobj & \ptpsub \\
        Transcription\cite{kong2021high} & \transobj & \transsub \\
        \midrule
        Proposed (PiCoGen2) & \ourobj & \oursub  \\
         - Ablation 1 (w/o pre-training) & \wopreobj & \wopresub \\
         - Ablation 2 (w/o fine-tuning) & \wofineobj & \wofinesub \\
        
        \midrule
        Human & \humanobj & \humansub \\
        \bottomrule
    \end{tabular}
    }
    \caption{
    The results of objective evaluations and the MOS of the subjective study ($\uparrow$\,/\,$\downarrow$: the higher\,/\,lower the better).
    }
    \label{table:mos}
\end{table*}

\subsection{Experiment Setup}

We implement PiCoGen2 using GPT-NeoX\cite{gpt-neox-library} as the piano token decoder and SheetSage\cite{donahue2022melody} cascaded with an adapter network as the song audio encoder.
The decoder consists of 8 layers, each with 8 attention heads.
The adapter is a 4-layer Transformer encoder with 8 attention heads per layer.
Our full model has approximately 39M learnable parameters, not counting the SheetSage part for we use it as is with its parameters frozen.

There are 2 ablations compared in the experiment, both of them sharing the same architecture as our full model, but
one ablation (Ablation 1) is trained on song-to-piano data \emph{without pre-training}, and
the other ablation (Ablation 2) is trained on piano-only data (i.e., \emph{without fine-tuning}).
For baselines, besides Pop2Piano, we also include the piano transcription model by Kong \emph{et al.}\cite{kong2021high} to validate the effectiveness of the encoder component in our model.

We train the models with Adam optimizer, learning rate 1e$-$4, batch size 4 and segment length 1,024.
The full model is pre-trained for 100K steps on the piano-only data, and then fine-tuned for an additional 70K steps on the song-to-piano paired data.
Ablation 1 is trained from scratch for 100K steps directly on the paired data.
Ablation 2 is trained for 50K steps only on the piano-only data, without any exposure to the song-to-piano pairs.
During the fine-tuning stage for the full model, we tune the weighting factor $\alpha$ that controls the balance between the piano-only loss and song-to-piano loss, and find that the model achieved the best performance when $\alpha$ is set to 0.25.



For the objective and subjective evaluations, all models are used to generate piano covers of the 95 testing songs (cf. Section \ref{db}).
To eliminate the bias caused by the varying quality of piano recordings, the ground truth human piano performances are first transcribed into MIDI note sequences. These MIDI sequences are then synthesized back into audio using the same FluidSynth-based MIDI synthesizer  \cite{fluidsynth} employed for the model outputs.

\subsection{Objective Metrics}
We adopt the following existing metrics to assess the quality of the generated piano covers from different aspects, including similarity to the original song and coherence of the piano performance itself.
\begin{itemize}[leftmargin=*,itemsep=0pt,topsep=2pt]
\item\textbf{Melody Chroma Accuracy ($MCA$)} \cite{raffel2014mir_eval} evaluates the similarity between two monophonic melody 
sequences.
The melody line plays a crucial role in deciding whether a song cover 
resembles
the original song.
Following Pop2Piano\cite{choi2023pop2piano}, we compute the MCA between the vocals extracted by Spleeter \cite{hennequin2020spleeter} from the test song audio, and the top melodic line extracted from the generated piano cover MIDI using the skyline algorithm \cite{skyline}. 

\item\textbf{Pitch Class histogram Entropy ($H_4$)} \cite{wu2020jazz} evaluates the harmonic diversity of a musical segment by computing the entropy of the distribution of note pitch class counts.
A lower entropy value indicates lower harmonic diversity, but implies a more stable and consistent tonality across the segment.
The subscript (``4'') indicates the number of bars over which the entropy is calculated. 


\item\textbf{Next-Bar Grooving Pattern Similarity ($GS$)} is modified from the grooving pattern similarity proposed in \cite{wu2020jazz}.
It originally measures the global rhythmic stability across an entire song. Instead of calculating over all pairs in the target, we adapt the metric to focus on local rhythmic stability within a song,
evaluating the rhythmic coherence between each bar and its succeeding bar.

\end{itemize}

\subsection{User Study}
For subjective evaluation, we conduct an online listening test involving 52 volunteers: 5 professional music producers, 13 amateurs, and 34 pro-amateurs with more than 3-year music training.
The volunteers are randomly assigned to distinct test sets, with each set containing 3 songs randomly selected from different genres, 
and for each song, there are 6 piano performances presented anonymously in random order.
These piano performances include: a human piano performance, outputs of our full model and the two ablated versions, and outputs of the Pop2Piano and piano transcription model baselines.
All of them are truncated to 40-second audio clips from the beginning.
Subjects are asked to listen to these audio clips and provide ratings on a 5-point Likert scale for the following aspects:

\begin{itemize}[leftmargin=*,itemsep=0pt,topsep=2pt]
  \item \textbf{Similarity (SI):} The degree of similarity between the piano performances and the original song.
  \item \textbf{Music Fluency (FL):} The degree of perceived fluency in the music, representing the smoothness and coherence of the piano performances.
  \item \textbf{Overall (OVL):} How much do the participants like the piano cover in the personal overall listening experience?
\end{itemize}

\subsection{Results}

\begin{figure*}
    \centerline{{
    \includegraphics[width=2.0\columnwidth]{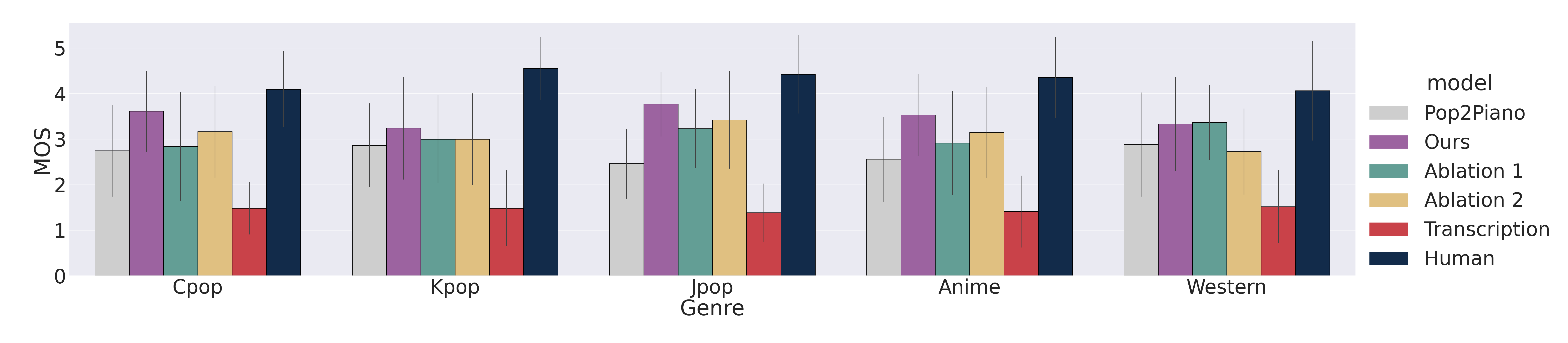}}}
    \vspace{-5mm}
    \caption{The MOS in overall scores (\textbf{OVL}) of the user study in different genres.}
    \label{fig:mos}
\end{figure*}

\tabref{table:mos} displays the results of the objective evaluation metrics and mean opinion scores (MOS) from the user study.
In the objective evaluation, Pop2Piano shows a leading MCA score compared to other models and the human piano performances,
which indicates it excels at matching the original song's melodic contour.
Except for the transcription baseline model, there is no significant difference in $GS$ and $H_4$ across models, suggesting comparable local rhythmic coherence and harmonic variety.



Next, we pay attention to the result of user study. 
Much to our delight, the full  model leads with the best scores across all aspects in the user study with statistical significance ($p < 0.05$), but there remains a gap compared to the human reference performances.
Ablation 1 achieves higher scores than Pop2Piano in all aspects of the user study, both of which are trained on the paired data.
This suggests that utilizing the weakly-aligned paired data, which avoids distorting the original piano performances, helps increase the overall listening experience quality of the model outputs for human raters.
Moreover, both Ablation 2 and the transcription baseline are trained on piano-only data, but Ablation 2 performs significantly better than the baseline in both objective and subjective evaluations.
This can be seen as evidence that SheetSage, as the encoder, extracts more relevant features beneficial for the piano cover generation task compared to the baseline transcription model.


\section{Discussion}



In the experiment, we note that while Pop2Piano exhibits a significantly higher MCA score than the other models, even higher than the human performances, it fails to achieve comparably high $SI$ ratings in the user study.
We suggest this conflict arises from the assumption in MCA that two melodies must temporally correspond to each other on a fixed ``time grid.'' That is, the corresponding chroma features must be located at precisely the same time instants.
For human listening experiences, two similar melodies only need to be coordinated on beats rather than a rigid time grid. Specifically, human perception of melodic similarity allows for the tempo or duration to be slightly changed in the same ratio, as long as their notes are located on the same underlying musical beat positions.
As mentioned in  Sections \ref{sec:introduction} \& \ref{sec:background}, different from transcription or arrangement, a cover song is not usually temporally aligned to the original song, i.e., the musical elements such as tempo, melody, rhythmic changed in the composition process of the piano cover.
This temporal flexibility suggests that MCA as an objective measure for the cover generation task may not be adequate and calls for future endeavor to develop better alternatives.

\begin{figure}
    \centerline{{
    \includegraphics[width=1.0\columnwidth]{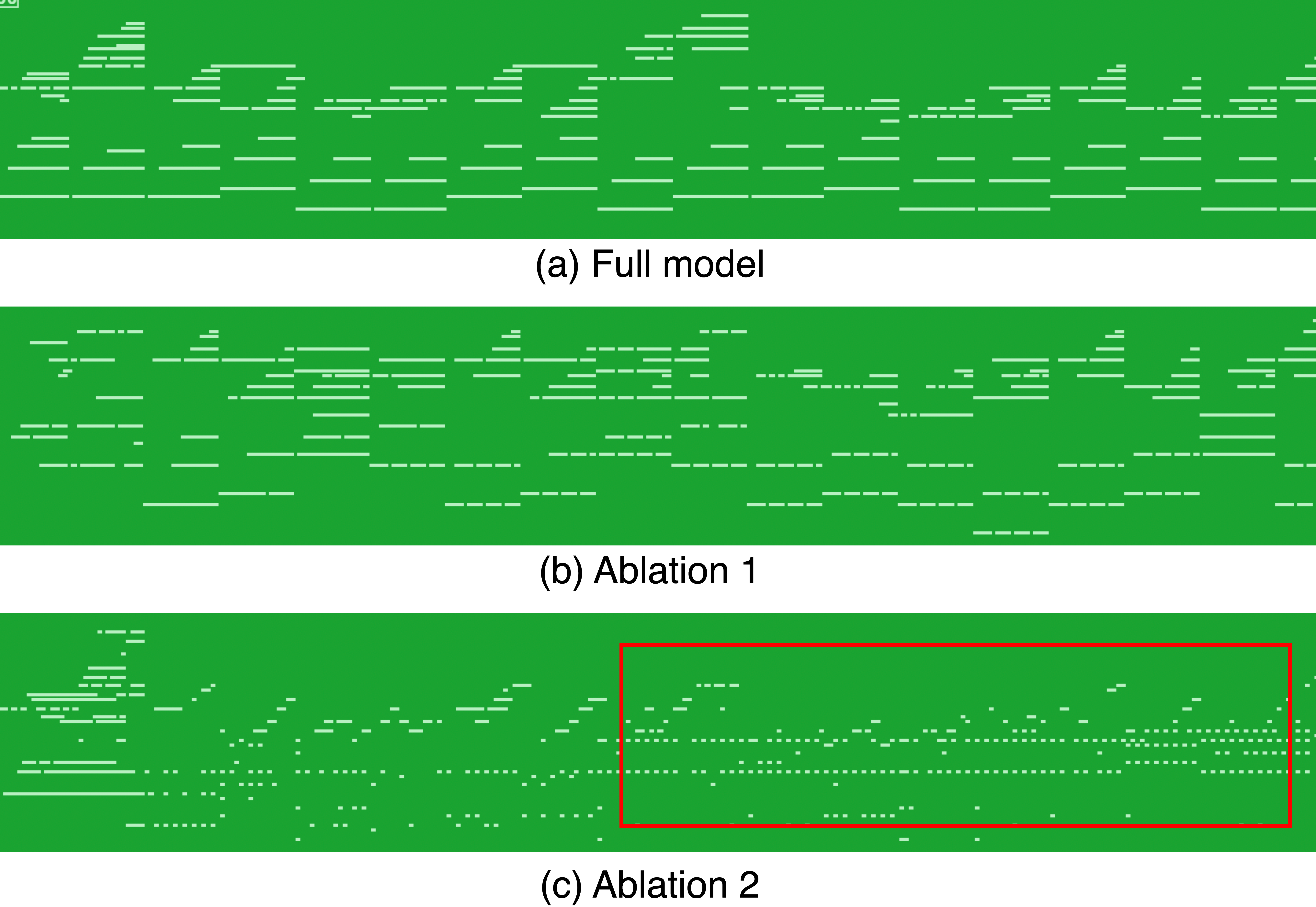}}}
    \vspace{-2mm}
    \caption{The pianoroll representation of a snippet from an example generated by the models. We observe that Ablation 2, which trained on piano-only data, tends to generate repeated short notes.}
    \label{fig:example}
\end{figure}

We also find that the two ablated models have the same \textbf{OVL} scores in the subjective evaluation, even though Ablation 2 has never seen any pop song data during training.
To investigate the reason behind this, we first examine the piano covers generated by Ablation 2.
\figref{fig:example} shows a snippet of a cover generated by this model.
We note that it tends to generate repeated short notes, resulting in an unnatural-sounding performance.
However, \figref{fig:mos} demonstrates the \textbf{OVL} scores across different music genres.
Interestingly, we see that Ablation 2 outperforms Ablation 1 for the \emph{Cpop}, \emph{Jpop}, and \emph{Anime} genres. Additionally, as shown in \tabref{table:mos}, the former ablation also achieves higher \textbf{SI} and lower \textbf{FL} scores than the latter.
From this observation, we suggest that (i) for short audio clips (less than 40 seconds), human raters may place more emphasis on initial melodic accuracy when judging the overall perceived quality, even if Ablation 1 tends to generate more coherent and natural-sounding results;
(ii) Ablation 1 does not effectively learn to precisely capture the melodic contour from the reference song condition due to the inherent alignment errors present in the weakly-aligned song-to-piano paired data it was trained on.


\section{Conclusion}

In this paper, we have presented PiCoGen2, which
applies the concept of transfer learning to the   piano cover generation task.
We propose a training strategy that involves two stages: pre-training on piano-only data to learn fundamental piano performance skills, followed by fine-tuning on weakly-aligned song-to-piano paired examples for the cross-domain translation.
A comprehensive set of experiments 
validate the effectiveness of the proposed transfer learning approach and the use of weakly-aligned data.



As we still require weakly-aligned data, future work can be done to tackle cover generation without relying on data alignment at all.
Moreover, it is useful to have a systematic analysis to evaluate the quality of piano covers and identify the key factors influencing the result, e.g., by studying the performance difference between PiCoGen~\cite{tan2024picogen} and PiCoGen2. It is also interesting to generate other covers, such as orchestral covers, and to develop better objective metrics.






\section{Acknowledgment}

The work is supported by a grant from the National Science and Technology Council of Taiwan (NSTC 112-2222-E-002-005-MY2). We are also grateful to the reviewers and meta-reviewer for helpful comments that help improve the quality of the paper.


\bibliography{ISMIRtemplate}

%
%
%
%
%

\end{document}